# Simple model for the RF field amplitude dependence of the trapped flux sensitivity in superconducting RF cavities

**Authors**

Sergio Calatroni, CERN, 1211 Geneva 23, Switzerland.
Ruggero Vaglio, Dipartimento di Fisica, Università di Napoli Federico II, CNR-SPIN e INFN, Napoli, Italy

**Abstract**

The improvement of the performance of superconducting RF cavities has recently motivated a considerable research effort in order to elucidate the effect of trapped magnetic flux on the surface resistance $R_s$. In this paper we show that by introducing a non-linear pinning force in the Gittleman-Rosenblum equations for the RF power dissipation due to a trapped magnetic flux in a superconductor, we can empirically describe the linear dependence on the RF field amplitude $B_{rf0}$ of the additional surface resistance $R_{fl}$. We also show that the proportionality between the RF-field dependent and independent terms $R_{fl}^1$ and $R_{fl}^0$, and the frequency dependence of $R_{fl}^1$ follow naturally from this approach.

## 1. Introduction

The quest for optimizing energy consumption in particle accelerators is motivating the research for increasing the quality factor in Superconducting Radio-Frequency (SRF) cavities, as underlined in recent projects [1], [2]. This in turn has motivated a renewed interest in several Laboratories to study the effect of trapped magnetic flux on SRF cavity performance, and on the possibilities of minimizing its consequences. The effect of trapped magnetic flux on the quality factor $Q = \Gamma/R_s$, where $R_s$ is the surface resistance and $\Gamma$ depends only on cavity geometry, has been experimentally identified and studied since the earliest developments of SRF cavities [3], [4], [5]. The experimental data are generally well described, in particular at low RF field amplitude, by an extra additive term $R_{fl}$ to the cavity surface resistance $R_s$, having the bi-linear form:

$$R_{fl}(B_{rf0}, B_0) = \left( R_{fl}^0 + R_{fl}^1 B_{rf0} \right) B_0, \tag{1}$$

where $B_{rf0}$ is the RF field amplitude, $B_0$ is the trapped external field, $R_{fl}^0$ is the zero RF field trapped flux sensitivity and $R_{fl}^1$ is the RF slope of the sensitivity. The first term is typically dominating in the case of bulk niobium cavities, although the second term can become of a comparable order of magnitude at high RF fields, in particular for high-frequency cavities [6], [7]. Interestingly, it was found that cavities made with the sputter thin film technology of niobium on a copper substrate (Nb/Cu) were much less sensitive to the ambient magnetic field, usually because of a much smaller $R_{fl}^0$ compared to bulk Nb cavities [5], [6]. The Earth magnetic field is large enough to induce a sensible degradation of the surface resistance and this motivated the use of efficient magnetic shielding of bulk Nb SRF cavities, and conversely Nb/Cu cavities need not be shielded in typical accelerator applications [8]. In recent years, it has also been realized that the expulsion or trapping of magnetic flux depends in large part on the cooldown dynamics and in particular on the temperature gradient along the cavity upon crossing the critical temperature [9], [10]. Although adequate cool-down conditions allow obtaining very large quality factors [11], understanding of the phenomenon has motivated several studies, in particular as a function of the niobium structure, treatment and mean free path [12], [13]. Recent investigations have also focused on the frequency dependence of the effect of trapped magnetic flux [7] confirming and integrating older results [3]. In this paper we will present a simple model based on a non-linear extension of the classical Gittelman and Rosemblum (GR) model [14], originally introduced to describe dissipations in thin films due to the motion of a rigid flux-lattice under the effect of an RF current as in a forced linear oscillator.

Our model reproduces the linear dependence of $R_{fl}$ on $B_{rf0}$ described by Eq. (1), and the experimental correlation found between $R_{fl}^0$ and $R_{fl}^1$. We will also show that the available experimental data on the frequency dependence of $R_{fl}^1$ can be well described by of our model.

## 2. The Gittelman and Rosemblum model

The original GR model considers a thin slab with uniform RF currents in its thickness $d$ and small displacements of a rigid vortex lattice due to the small applied RF current compared to the critical current. This last point corresponds to the assumption of a harmonic potential $U(x) = \frac{1}{2}kx^2$, where $x$ is the displacement from the equilibrium position and $k$ the elastic constant per unit fluxon length. This results in an equation of motion for a typical flux tube:

$$\eta \dot{x}(t) + k x(t) = \phi_o J_{rf0} \sin \omega t \tag{2}$$

Here $\eta = \frac{\phi_o B_{c2}}{\rho_n}$ is the flow viscosity per unit length with $\phi_0$ the flux quantum, $B_{c2}$ the superconductor upper critical field and $\rho_n$ the normal-state electrical resistivity, $J_{rf0}$ is the maximum amplitude of the RF current of angular frequency $\omega$, and $\phi_o J_{rfo}$ is the maximum force per unit length exerted by the RF field on the flux lines. The effective mass per unit length $m$ of the fluxons is small and the acceleration term $m\ddot{x}(t)$ can typically be neglected [14], [15], which simplifies the calculations. The flux lattice displacement is then:

$$x_o(t) = x_o \sin(\omega t - \varphi) \tag{3}$$

With:

$$x_0 = \frac{J_{rf0} \phi_0}{\eta \sqrt{\omega^2 + \omega_0^2}}; \quad tg\varphi = \frac{\omega}{\omega_0} \tag{4}$$

From the calculated fluxon velocity, it is straightforward to calculate the average power dissipated by the fluxon motion. The surface resistance due to flux motion is then obtained by the standard relation:

$$P_s = \frac{1}{2} R_{fl} H_{rfo}^2, \text{ with } H_{rf0} = J_{rf0}d = \frac{B_{rf0}}{\mu_0}. \tag{5}$$

The model can also be extended to the case of isolated vortices trapped by pinning centers of radius $r_0 < \xi$. In this case the pinning force per unit length is $k \cong \frac{J_c \phi_0}{\xi}$ with $J_c$ the depinning current density [16] and $\xi$ is the coherence length. The overall power dissipation is then obtained by summing the individual flux lines dissipation. The result is the same as in the original approach:

$$R_{fl} = R_n \frac{B_o}{B_{c2}} \frac{\omega^2}{\omega^2 + \omega_o^2} \quad , \qquad (6)$$

with $\omega_0 = \frac{k}{\eta}$, $B_0 = n\phi_0$ and $R_n$ is the normal-state surface resistance ($R_n = \frac{\rho_n}{d}$ for a thin slab of thickness $d$). Note that Eq. (6) in the limit $\omega \gg \omega_o$ gives $R_{fl}(0) = R_n \frac{B_o}{B_{c2}}$ which reproduces the result of the so-called "static model" [17]. The trapped flux sensitivity $S$ deduced from Eq. (6) is :

$$S = \frac{R_{fl}}{B_o} = \frac{R_n}{B_{c2}} \frac{\omega^2}{\omega^2 + \omega_o^2} \qquad (7)$$

A crucial parameter of the model is the depinning frequency $\omega_0$ that separates the "pinning regime" $\omega \ll \omega_0$ from the "flux flow regime" $\omega \gg \omega_0$ where the viscous force is larger than the elastic force resulting in a steep increase of the dissipated power. This gives a possible interpretation on why thin films cavities present a lower sensitivity to trapped flux in respect to the bulk. Indeed Nb thin films present a higher depinning frequency in respect to high RRR bulk materials so that the condition $\omega \ll \omega_0$ is satisfied at typical accelerating cavities frequencies [18].

Though the original GR model was considering very thin films, the model can be extended in principle to film thicknesses smaller than the pinning correlation length $L_c \cong 2\xi\sqrt{J_d/J_c}$ (here $J_d$ is the depairing current density), that can be over 1 μm for Nb films [19].

Various extensions of the model have been recently proposed in the literature [20], [21] showing that the GR-type models can fit also the experimental results of pure bulk cavities, though the assumption of perpendicular rigid vortices driven by uniform RF currents seems to be hardly applicable to describe the long flexible vortices trapped in bulk Nb cavities. Starting from this last consideration, a model of non-rigid fluxons including both bending modes and the nonlocal tension of the vortex line has been recently introduced [22], describing experimental results with very good accuracy [23], [13].

It is important to underline here that all the models based on the GR simple picture describe the pinned flux lines under the effect of the RF currents as forced linear oscillators, so that the power dissipation is always proportional to the square of the RF field amplitude, hence the surface resistance does not depend on $B_{rf}$. In different words, these models can only describe the term $R_{fl}^0$ but do not provide a description of $R_{fl}^1$. In the following we will extend the basic GR model, introducing an ad hoc non-quadratic pinning potential (non-linear pinning force) in order to reproduce the linear $B_{rf}$ field amplitude dependence experimentally observed by many authors [3], [5], [6], [7]. We should mention that a recent different approach also attempts at explaining the dependence on $B_{rf}$, by making use of a mean-field collective pinning potential [24].

## 3. Non linear flux oscillation model

Let us assume a non-harmonic pinning potential:

$$U(x) = \frac{1}{2}kx^2 - \frac{1}{3}\gamma |x^3| \qquad (8)$$

This form of the pinning potential is a reasonable approximation of theoretical predictions of the real potential well shape [25]. The pinning force per unit length corresponding to this potential is:

$$f_p = -kx + \gamma |x| x \qquad (9)$$

$U(x)$ and $f_p(x)$ are plotted in Fig. 1 for different values of the adimensional parameter $\alpha = \dfrac{\gamma \bar{x}}{k}$

where $\bar{x}$ is the maximum vortex displacement from the equilibrium position and α represents the ratio of the nonlinear to the linear component of the pinning force at the maximum vortex displacement. It can be seen that for $\alpha > 0.5$, the force is non-monotonic in $x$ corresponding to a non-physical behavior.
In the same approximations of the GR model we need now to solve the non-linear differential equation:

$$\eta \dot{x}(t) + kx(t) - \gamma |x(t)| x(t) = \phi_0 J_{rf0} \sin \omega t \qquad (10)$$

Eq. (10) can be rewritten in terms of the adimensional parameter $\alpha$:

$$\frac{1}{\omega_o}\dot{x} + x - \alpha \left|\frac{x(t)}{\bar{x}}\right| x(t) = \frac{J_{rfo}\varphi_O}{k} \sin \omega t \qquad (11)$$

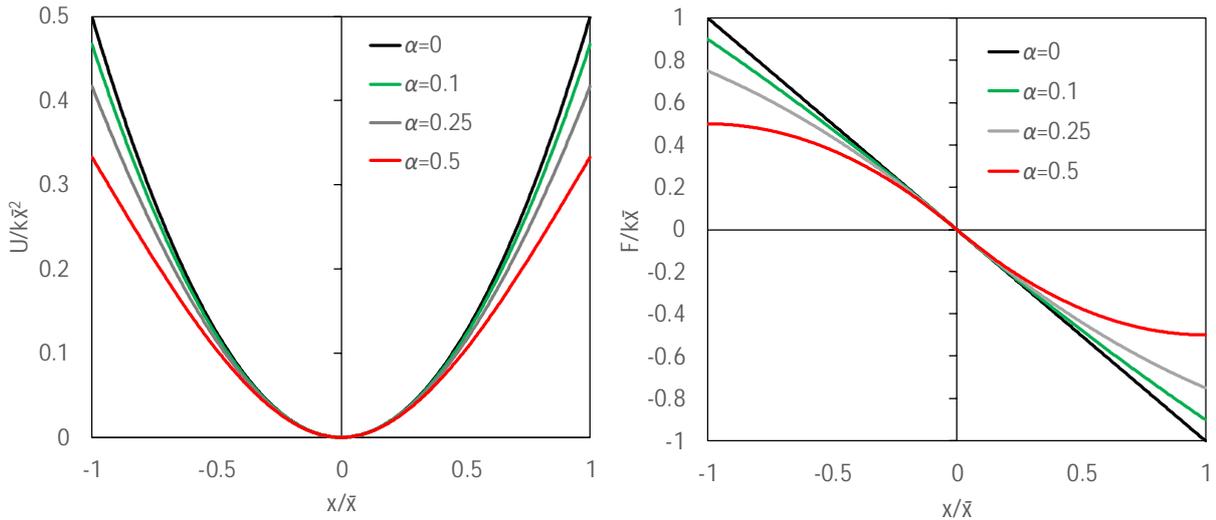

FIG. 1. Plots of $U(x)$ (left) and $f_p(x)$ (right) as a function of the relative vortex displacement $x/\bar{x}$ for various values of the parameter α.

For a small nonlinear term, i.e. for $\alpha \ll 1$, we can solve this equation by classical perturbation methods, expanding $x(t)$ in powers of $\alpha$. At the first order:

$$x(t) = x_o(t) + \alpha x_1(t) \tag{12}$$

were $x_o(t) = x_o \sin(\omega t - \varphi)$ is the same solution as for the linear case ($\alpha = 0$), with $x_o$ and $\varphi$ already defined by Eq. (4).

By substituting Eq. (12) in Eq. (11) and neglecting the terms above the first order in $\alpha$, we have the linear differential equation for $x_1(t)$:

$$\frac{1}{\omega_o}\dot{x}_1(t) + x_1(t) = \left|\frac{x_o(t)}{x_o}\right| x_o(t) \tag{13}$$

Where at the same approximation order we can write $\alpha = \frac{\gamma x_o}{k}$ since $\alpha = \frac{\gamma \overline{x}}{k} = \frac{\gamma x_o}{k} + \alpha \frac{\gamma \overline{x}_1}{k}$.

The instant power dissipation due to the single vortex line flow motion is $p_{sfl}(t) = d\, f_L(t) v(t)$ where $f_L(t) = \varphi_o J_{rf0} \sin \omega t$ is the force per unit length and $v(t) = \dot{x}(t) = \dot{x}_o(t) + \alpha \dot{x}_1(t)$ the vortex velocity, and to get the power loss we have to multiply for the slab thickness $d$ (as in the GR approach the force is assumed constant along the vortex length). We can now calculate the average power per unit surface $P_{sfl}$ dissipated per cycle by a single fluxon as:

$$P_{sfl} = P_{sflo} + P_{sfl1} = d\, \phi_o J_{rfo} \frac{1}{T}\int_0^T \dot{x}_o(t)\sin\omega t\, dt + d\, \phi_o J_{rfo} \alpha \frac{1}{T}\int_0^T \dot{x}_1(t)\sin\omega t\, dt \tag{14}$$

Here, obviously, $P_{sflo}$ corresponds to the single fluxon power dissipation in the linear GR model.

Expanding $x_1(t)$ in Fourier series:

$$x_1(t) = \sum_{-\infty}^{+\infty} A_n e^{in\omega t} \;;\; A_n = \frac{1}{T}\int_0^T x_1(t) e^{-in\omega t} dt \tag{15}$$

Integration by parts of the second integral in Eq. (14) shows that:

$$P_{sfl1} = -d\phi_o J_{rfo} \alpha \omega \operatorname{Re}[A_1] \tag{16}$$

Integration of Eq. (13) then gives:

$$(1 + i\frac{\omega}{\omega_o})A_1 = \frac{1}{T}\int_0^T \left|\frac{x_o(t)}{x_o}\right| x_o(t) e^{-i\omega t} dt \tag{17}$$

Which results in:

$$\text{Re}[A_1] = \frac{\omega_o x_o}{\omega^2 + \omega_o^2} \frac{1}{T} \int_0^T (\omega_o \cos \omega t - \omega \sin \omega t) \left| \frac{x_o(t)}{x_o} \right| \frac{x_o(t)}{x_o} dt \qquad (18)$$

The calculation of the integrals in Eqs. (14) and (18) is drafted in the Appendix. Combining the result of Eq. (A6) with the expressions of Eq. (5) we finally obtain:

$$R_{fl}(B_{rfo}) = R_{fl}(0) \left[ 1 + \frac{16\alpha}{3\pi} \frac{\omega_0^2}{(\omega^2 + \omega_o^2)} \right] \qquad (19)$$

with:

$$R_{fl}(0) = R_n \frac{B_o}{B_{c2}} \frac{\omega^2}{\omega^2 + \omega_o^2} \qquad (20)$$

The parameter $\alpha$ contains the dependence on $B_{rf0}$:

$$\alpha = \frac{\gamma x_o}{k} = \frac{\gamma \phi_o J_{rfo}}{k\eta \sqrt{\omega^2 + \omega_o^2}} = \frac{\gamma}{k} \frac{R_n}{\mu_o} \frac{B_{rfo}}{B_{c2}} \frac{1}{\sqrt{\omega^2 + \omega_o^2}} \qquad (21)$$

For $\alpha=0$, i.e. for $\gamma=0$ or $B_{rfo}=0$ the GR result of Eq. (6) is reproduced.
Our nonlinear model also reproduces the phenomenological Eq. (1) with:

$$R_{fl}^0 = \frac{R_n}{B_{c2}} \frac{\omega^2}{\omega^2 + \omega_o^2} \qquad (22)$$

$$R_{fl}^1 = R_{fl}^o \frac{16}{3\pi} \frac{\gamma}{k} \frac{R_n}{\mu_o B_{c2}} \frac{1}{\sqrt{\omega^2 + \omega_o^2}} \frac{\omega_o^2}{(\omega^2 + \omega_o^2)} \qquad (23)$$

TABLE I. Coefficients $R_{fl}^0$ and $R_{fl}^1$ as measured by several authors. The first lines are for bulk Nb, while the bottom lines are for Nb/Cu. (*) Recalculated from data reported in the original paper for measurements of helically loaded "Helix II" resonators at various harmonics. (**) Data extracted from the published plots for "120 °C baking". Different frequencies correspond to different elliptical cavities. (***) The data correspond to the A/B families identified in the original paper ("oxidized Cu" / "oxide-free Cu").

| Author | Frequency [MHz] | $R_{fl}^0$ [nΩ/G] | $R_{fl}^1$ [nΩ/(G·mT)] |
| --- | --- | --- | --- |
| Piosczyk [3] (*) | 91 / 160 / 290 | 3.5 / 9.5 / 28 | 0.35 / 0.55 / 0.9 |
| Arnolds-Mayer [5] | 500 | 150 | 5 |
| Checchin [7] (**) | 650 / 1300 / 2600 / 3900 | 700 / 1000 / 1500 / 1900 | 1.6 / 2.6 / 6.1 / 7.4 |
| Miyazaki [26] | 101 | 3.2 | 0.32 |
| Benvenuti [27] (***) | 1500 | 3.3 / 56 | 0.91 / 4.5 |

## 4. Comparison with experimental results

The model presented in the previous section predicts a linear dependence of the flux sensitivity $S = R_{fl}/B_0$ on the amplitude of the RF field $B_{rf0}$. This is indeed observed in several experiments, both on bulk Nb cavities [3], [5], [7] and Nb/Cu thin film cavities [6], [26], [27], and allows expressing the flux sensitivity as in Eq. (1). The coefficients of flux sensitivity and RF field dependence measured in these experiments are reported in Table 1.

In practice, however, this is a somewhat trivial result, since the shape of the non-quadratic pinning potential (non-linear pinning force) has also been selected to reproduce this result. A higher order of the potential would have naturally resulted in a higher-order RF field dependence.

From the experimental data it is always possible to deduce the value of the non-linear parameter $\alpha$ and verify whether this term is small enough to be compatible with the adopted perturbative approach. From Eq. (19) and in the hypothesis $\omega < \omega_o$, certainly adequate for Nb films, the maximum RF field $B_{rf0}^{max}$ for which the minimum condition $\alpha < 0.5$ is satisfied corresponds to $R_{fl}(B_{rf0}^{max}) \lesssim 2R_{fl}(0)$. As an example, for the Nb/Cu films reported in Fig. 2 this is true up to a $B_{rf0} \leq 5mT$, however the range of the linear dependence of the flux sensitivity on $B_{rf0}$ extends much beyond this value. We should nevertheless underline that at the largest amplitudes the flux sensitivity deviates from linearity.

Our non-linear model predicts also other non-trivial experimental results. Indeed Eq. (23) shows a proportionality between $R_{fl}^o$ and $R_{fl}^1$ as observed on bulk Nb [3], [6], on Nb/Cu thin films [6], [27], and also on Nb$_3$Sn films grown by thermal diffusion of Sn vapours inside Nb cavities [28]. Data extracted from [6] for several Nb/Cu cavities coated with different sputter gases (Ar, Kr and Xe) produced at CERN are reported in Fig. 3, which illustrates well the predicted behavior.

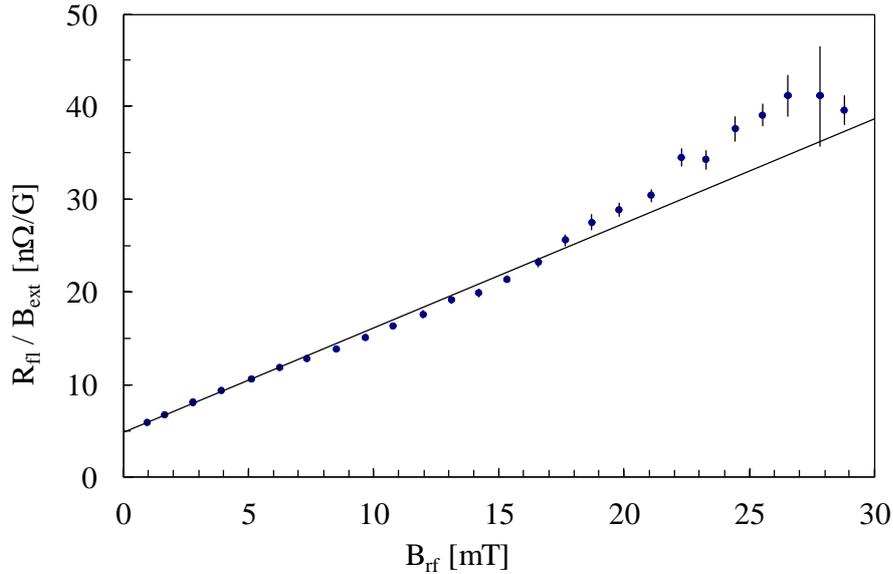

FIG. 2. The dependence on $B_{rf}$ of $R_{fl}(1.7K)/B_{ext}$ for Nb/Cu cavities. The line is a fit to the data of the form $R_{fl}^0 + R_{fl}^1 B_{rf}$. (Reprinted from [6] with STM permission).

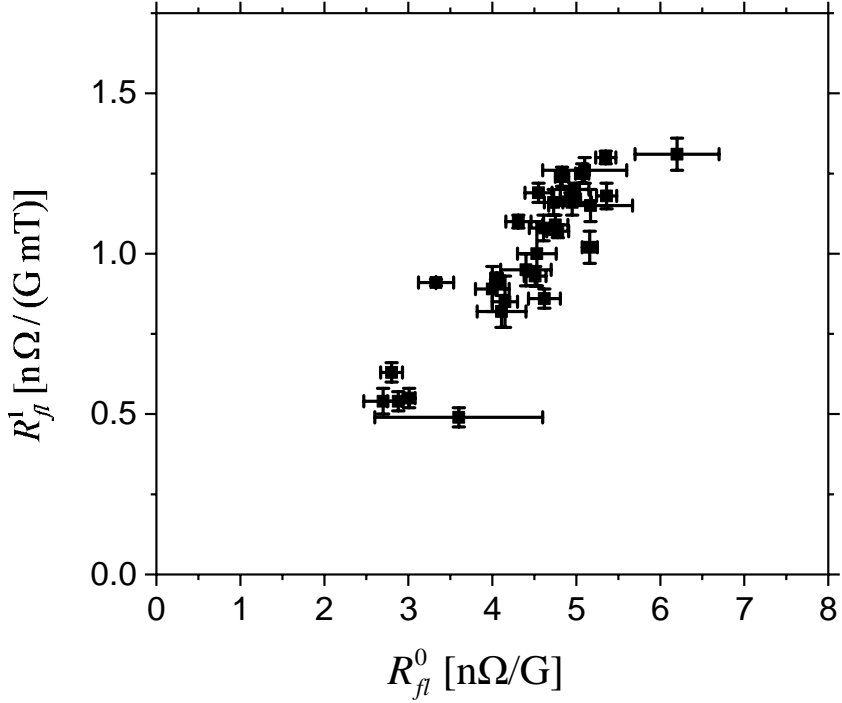

FIG. 3. Correlation plot of $R_{fl}^o$ and $R_{fl}^1$ for Nb/Cu films at 1500 MHz [6].

The present model can also describe the frequency-dependence of the effect of trapped flux. From Eqs. (22) and (23) we can write:

$$R_{fl}^1 \propto \frac{1}{\sqrt{\omega^2 + \omega_o^2}} \left( \frac{\omega_o \omega}{\omega^2 + \omega_o^2} \right)^2 \qquad (24)$$

This dependence is plotted in Fig. 4 for $\omega \leq \omega_o$, fitted to the available data from recent experiments on "120 °C baked" bulk Nb cavities [7], and to older experimental results at lower frequency which date back to the very early developments of SRF using low-RRR Nb [3], also reported in the figure.

One should be aware, however, that Eq. (24) predicts a decrease of $R_{fl}^1$ vs $\omega$ for $\omega \geq \omega_o$ not observed in the experiments.

## 5. Conclusions

We have shown that, by inserting a simple non-harmonic pinning potential in the GR equations, we can correctly reproduce the linear dependence on $B_{rfo}$ of the losses induced by an external trapped magnetic field in SRF cavities. A proportionality between $R_{fl}^o$ and $R_{fl}^1$ follows naturally from this approach. In the limit $\omega < \omega_0$, the frequency dependence of $R_{fl}^1$ observed in the experiments can also be reproduced.

We are aware that our simple model cannot reproduce all the large body of experimental evidence, however its good capability of describing some of the main features of the flux sensitivity could be the basis for a more complete theoretical treatment based on more refined models, which would include the specific mechanisms producing the non-linear pinning force.

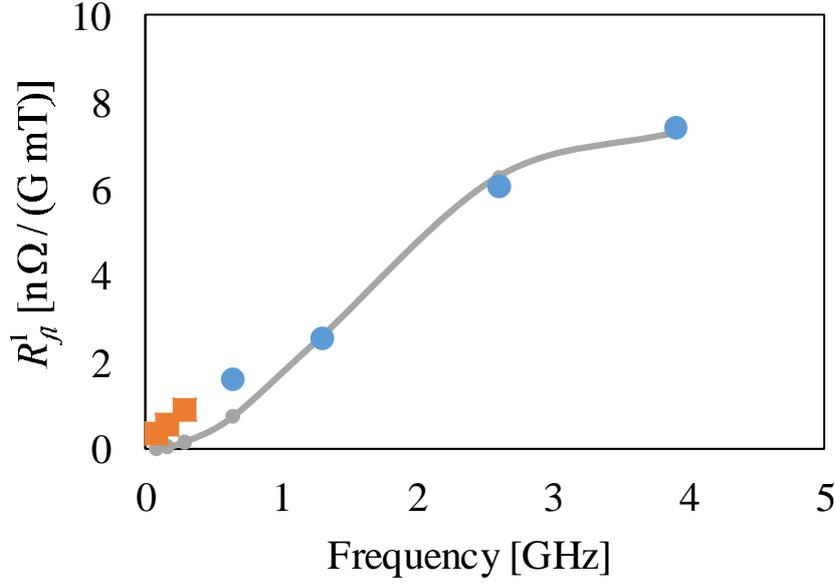

FIG. 4. Dependence of $R_{fl}^1$ on frequency. Data on "Helix-II" low-frequency resonators from [3] (squares) and on "120 °C baked" elliptical cavities from [7] (circles). The gray line is a best fit to the data, resulting in a depinning frequency $\nu_0 = \omega_0 / 2\pi = 4.6$ GHz

### Acknowledgements

The authors would like to acknowledge Alex Gurevich (Old Dominion University) for his helpful suggestions on the techniques for solving non-linear equations and the valuable comments on the manuscript. The authors would also like to acknowledge several stimulating discussions with Akira Miyazaki and Walter Venturini Delsolaro (CERN). R. V. has been supported by the INFN V Group Experiment ISIDE

### Appendix

From Eqs. (14) and (18) we have: (A1)

$$P_{sfl} = P_{sfl0} + P_{sfl1} = d\,\phi_o J_{rfo}\omega x_o \cdot$$

$$\left\{ \frac{1}{T}\int_0^T \cos(\omega t - \varphi)\sin\omega t\, dt + \frac{\alpha\omega_o}{(\omega^2 + \omega_o^2)}\left[\omega\frac{1}{T}\int_0^T \sin\omega t\,|\sin(\omega t-\varphi)|\sin(\omega t-\varphi)dt - \omega_o\int_0^T \cos\omega t\,|\sin(\omega t-\varphi)|\sin(\omega t-\varphi)dt\right]\right\}$$

The integrals in Eq. (A1) can easily be computed giving respectively:

$$\frac{1}{T}\int_0^T \cos(\omega t - \varphi)\sin\omega t\, dt = \frac{1}{2}\sin\varphi \qquad (A2)$$

$$\frac{1}{T}\int_0^T \sin\omega t\,|\sin(\omega t - \varphi)|\sin(\omega t - \varphi)dt = \frac{4}{3\pi}\cos\varphi \qquad (A3)$$

$$\frac{1}{T}\int_0^T \cos\omega t \, |\sin(\omega t - \varphi)| \sin(\omega t - \varphi) dt = -\frac{4}{3\pi}\sin\varphi \tag{A4}$$

Then, using the result of Eq. (4) for $x_o$ and $\varphi$, we get:

$$P_{sfl} = \frac{1}{2}\frac{d\,\phi_o^2 J_{rfo}^2}{\eta}\frac{\omega^2}{\omega^2+\omega_o^2}\left[1+\frac{16}{3\pi}\alpha\frac{\omega_o^2}{\left(\omega^2+\omega_o^2\right)}\right] \tag{A5}$$

If we have $n$ vortices per unit surface characterized by the same depinning current, since the overall power dissipated per unit surface is obviously $P_s = nP_{sfl}$ and remembering that $B_0 = n\phi_0$, $H_{rfo} = dJ_{rfo}$, $\eta = \frac{\phi_o B_{c2}}{\rho_n}$, and $R_n = \frac{\rho_n}{d}$ (for a thin slab), we finally obtain:

$$P_s = \frac{1}{2}R_n\frac{B_o}{B_{c2}}\frac{\omega^2}{\omega^2+\omega_o^2}\left[1+\frac{16\alpha}{3\pi}\frac{\omega_0^2}{\left(\omega^2+\omega_o^2\right)}\right]H_{rfo}^2. \tag{A6}$$